\documentclass{kluwer}    

\newdisplay{guess}{Conjecture}

\def\Frac#1#2{{{\displaystyle\strut#1}\over{\displaystyle\strut#2}}}

\def\crm{\cr\noalign{\medskip}}
\def \be  {\begin{equation}}
\def \ee  {\end{equation}}

\def\EQM#1{\vcenter{\normalbaselines{\mathsurround=0pt}
    \ialign{${\displaystyle ##}$\hfil&&\ ${\displaystyle ##}$\hfil\crcr
    \mathstrut\crcr\noalign{\kern-\baselineskip}
    \noalign{\smallskip}
    #1\crcr\mathstrut\crcr\noalign{\kern-\baselineskip}}}}

\def\etal{{\it et al.}}
\def\llabel#1{\label{#1}} 
\def\g{\gamma}
\def\a{\alpha}
\def\e{{\rm \bf e}}

\begin{document}                                                                                   
\begin{article}
\begin{opening}         
\title{Note on the generalized Hansen and Laplace coefficients} 
\author{Jacques Laskar}  
\runningauthor{Jacques Laskar}
\runningtitle{Note on the generalized Hansen coefficients}
\institute{Astronomie et Syst\`emes Dynamiques, IMCCE-CNRS UMR8028, 
Observatoire de Paris,
77 Av. Denfert-Rochereau, 75014 Paris, France}
\date{\today}

\begin{abstract}
Recently, Breiter \etal (2004) reported the  computation of Hansen coefficients 
$X_k^{\gamma,m}$ for non integer values of $\gamma$. In fact,
the  Hansen coefficients are closely related to the Laplace $b_{s}^{(m)}$, and 
generalized Laplace coefficients
$b_{s,r}^{(m)}$ (Laskar and Robutel, 1995) that do not require $s,r$ to be integers.
In particular, the coefficients $X_0^{\g,m}$  have very simple expressions in terms 
of the usual Laplace coefficients $b_{\g+2}^{(m)}$, and all their 
properties derive easily from the known properties of the Laplace coefficients.
\end{abstract}
\end{opening}           

\section{Introduction}
The aim of this note is to clarify some simple relations between 
the Hansen coefficients, and the Laplace's, and generalized Laplace coefficients.
Once these relations are explicited, the results quoted by Breiter \etal (2004)
become simple translations of known results on the Laplace coefficients.

\section{Hansen and Laplace coefficients}  
The Hansen coefficients (Hansen, 1855)  are defined as the Fourier coefficients $X_k^{\g,m}$ 
of the series
\be
\left(\Frac{r}{a}\right)^\g \e^{i m v} = \sum_{k=-\infty}^{+\infty}  X_k^{\g,m} \e^{ikM}
\ee
where $v,M$ are the true and mean anomaly, $r,a$ the radial distance and semi-major axis. 
The transformation $v\rightarrow -v$ transforms $M$ in $-M$. Thus $X_k^{\g,m}$ is real
and $X_{-k}^{\g,-m}=X_k^{\g,m}$. We have  
\be
X_k^{\g,m} = \frac{1}{2\pi}\int_0^{2\pi} \left(\Frac{r}{a}\right)^\g \e^{ i m v} \e^{ -ikM} \, dM \ .
\llabel{eq.hans}
\ee

In particular, for $k=0$,
\be
X_0^{\g,m} = \Frac{1}{\sqrt{1-e^2}}\Frac{1}{2\pi}\int_0^{2\pi} \left(\Frac{r}{a}\right)^{\g+2} \e^{ i m v} \, dv \ .
\ee
Hansen (1855) uses the  expressions in term of the true anomaly $v$  
\be
\Frac{r}{a} = \Frac{1-e^2}{1+e\cos v} = \Frac{(1-e^2)(1+\beta^2)}{(1+\beta\xi)(1+\beta\xi^{-1})}
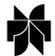
\ee 
where  $\xi = \e^{iv}$, and  $\beta = (1-\sqrt{1-e^2})/e$. We thus obtain immediately the expansion 
of $(r/a)^{\g+2}$ in Laurent series of $\xi$
\be
\left(\Frac{r}{a}\right)^{\g+2} = (1-e^2)^{\g+2}(1+\beta^2)^{\g+2} 
\Frac{1}{2} \sum_{k=-\infty}^{+\infty} b_{\g+2}^{(k)}(-\beta) \xi^k 
\label{eq.5}
\ee
where $b_s^{k}(\alpha)$ are the classical Laplace coefficients defined as the coefficients of the Laurent series
\be
(1-\alpha z)^{-s} (1-\alpha z^{-1})^{-s}=  \Frac{1}{2}   \sum_{k=-\infty}^{+\infty} b_{s}^{(k)}(\alpha) z^k  \ ,
\ee
with $b_s^{(k)}(-\alpha)= (-1)^k b_s^{(k)}(\alpha)$;   $b_s^{(-k)}(\alpha)= b_s^{(k)}(\alpha)$, and for $k \geq 0$,
\be
b_s^{(k)}(\alpha) = \Frac{(s)_k}{k!}\alpha^k F(s,s+k,k+1; \alpha^2) \ ,
\ee
where $(s)_0=1$, $(s)_k= s(s+1)\cdots(s+k-1)$ for $k \geq 1$. Thus
\be
X_0^{\g,m} = \Frac{(-1)^m}{2} (1-e^2)^{\g+3/2}(1+\beta^2)^{\g+2} b_{\g+2}^{(m)}(\beta) \ .
\ee

Any property of  the Laplace coefficients can thus be translated into a property on the
Hansen coefficients $X_0^{\g,m}$. 
In particular, in 1785, Laplace demonstrated the most 
useful relations 
\be
\EQM{
 b_{s+1}^{(j)}(\a)  &= \Frac{(s+j)}{s}\Frac{(1+\a^2)}{(1-\a^2)^2} b_s^{(j)}(\a)   
                  -\Frac{2(j-s+1)}{s} \Frac{\a }{(1-\a^2)^2} b_s^{(j+1)}(\a) \crm
b_{s+1}^{(j+1)}(\a)  &=  \Frac{j}{j-s}(\a+\Frac{1}{\a})   b_{s+1}^{(j)}(\a)   
                   - \Frac{j+s}{j-s} b_{s+1}^{(j-1)}(\a)
}
\ee
that are immediately translated as\footnote{Eq. \ref{eq.rec2}
 is the same as Eq. (19) of (Breiter \etal, 2004).}
\be
X_0^{\g,m} = \Frac{\g+1+m}{\g+1} X_0^{\g-1,m} + \Frac{m-\g}{\g+1} e X_0^{\g-1,m+1} 
\llabel{eq.rec1}
\ee
\be
X_0^{\g,m+1} = -\Frac{2}{e}\Frac{m}{m-\g-1} X_0^{\g,m} - \Frac{m+\g+1}{m-\g-1} X_0^{\g,m-1}  \ .
\llabel{eq.rec2}
\ee

As with the Laplace coefficients, 
these relations allow to express all coefficients with 
respect to the two first ones $ X_0^{\g,0}$ and $ X_0^{\g,1}$. Breiter \etal (2004) treat as a special case 
$\g=(2n+1)/2$. This is precisely the case of the  expansion of the Newtonian potential
 in Laplace coefficients. In fact,  the recurrence formulas of Laplace (\ref{eq.rec1}, \ref{eq.rec2}), allow to 
express all coefficients not with 4 initial coefficients, as quoted in Breiter \etal, 2004,
 but from only two of them, namely
\be
X_0^{-3/2,0} = (1+\beta^2)^{1/2}\Frac{1}{2} b_{1/2}^{(0)}(\beta)\ ;\qquad 
X_0^{-3/2,1} = -(1+\beta^2)^{1/2}\Frac{1}{2} b_{1/2}^{(1)}(\beta)\ ,
\ee
with the expressions in terms of the elliptic integrals of first and second kind $K(\beta),E(\beta)$ (Tisserand, 1889)
\be
 b_{1/2}^{(0)}(\beta) = \Frac{4}{\pi} K(\beta) \ ;\qquad 
  b_{1/2}^{(1)}(\beta) = \Frac{4}{\pi\beta} (K(\beta)-E(\beta) ) \ .
\ee
More generally,  $ X_0^{\g,m}$ can be expressed with respect to $ X_0^{\g-[\g],0}$ and $ X_0^{\g-[\g],1}$, where
$[\g]$ denotes the integer part of $\g$.

\section{Hansen coefficients for $k\in \dZ$}

The expressions of the Hansen coefficients for $k\neq 0$ are  obtained in a similar way 
using expansions of (\ref{eq.hans}) in eccentric anomaly $E$ as
\be
X_k^{\g,m} = \Frac{1}{2\pi} \int_0^{2\pi} \left(\Frac{r}{a}\right)^{\g+1} \e^{imv} \e^{ike\sin E} \e^{-ikE} dE \ .
\ee
With $\eta=\e^{iE}$ and 
\be
\Frac{r}{a} =\Frac{1}{1+\beta^2} (1-\beta\eta)(1-\beta \eta^{-1}) ;\quad 
\e^{i v} = \eta\Frac{(1-\beta\eta^{-1})}{(1-\beta\eta)} \ ,
\ee
we have
\be
\left(\Frac{r}{a}\right)^{\g+1} \e^{imv} 
= \Frac{\eta^m}{2(1+\beta^2)^{\g+1}} \sum_{l=-\infty}^{+\infty} b_{-\g-1+m,-\g-1-m}^{(l)}(\beta) \eta^l
\label{eq.16a}
\ee
where $b_{s,r}^{(k)}$ are the generalized Laplace coefficients defined in (Laskar and Robutel, 1995) as
the coefficients of the Laurent series
\be
(1-\alpha z)^{-s} (1-\alpha z^{-1})^{-r}=  \Frac{1}{2}   \sum_{k=-\infty}^{+\infty} b_{s,r}^{(k)}(\alpha) z^k  \ .
\label{eq.16}
\ee
We have  $b_{s,r}^{(k)}(-\a)= (-1)^k b_{s,r}^{(k)}(\a)$;  $b_{s,r}^{(-k)}(\a)= b_{r,s}^{(k)}(\a)$, and    for $k\geq 0$  
\be
b_{s,r}^{(k)}(\alpha) = \Frac{(s)_k}{k!}\alpha^k F(r,s+k,k+1; \alpha^2) \ .
\ee
The classical expansion in Bessel functions
\be
\e^{ike\sin E} =\sum_{n=-\infty}^{+\infty} J_n(ke) \eta^n \ ,
\ee
 allows  the computation of the Hansen coefficients $X_k^{\g,m}$ in terms of Bessel functions and 
 Laplace coefficients   as
\be
X_k^{\g,m} = \Frac{1}{2(1+\beta^2)^{\g+1}} \sum_{n=-\infty}^{+\infty} b_{-\g-1+m,-\g-1-m}^{(k-n-m)}(\beta) J_n(ke) \ .
\label{eq.19}
\ee
\section{Recurrence relations}
As for the Laplace coefficients, one can derive recurrence relations for the 
generalized Laplace coefficients.
Multiplying Eq. (\ref{eq.16}) by $A=1-\a z $ and respectively $B= 1-\a z^{-1}$, one derives 
the two recurrence relations 
\be
b_{s,r+1}^{(k)}(\a) -\a b_{s,r+1}^{(k+1)}(\a) = b_{s,r}^{(k)}(\a) \ ;
\label{eq.21}
\ee
\be
b_{s+1,r}^{(k)}(\a) -\a b_{s+1,r}^{(k-1)}(\a) = b_{s,r}^{(k)}(\a)Ê\ .
\label{eq.22}
\ee
The derivative of (\ref{eq.16}) with respect to $z$ gives also
\be
\a s A^{-s-1}B^{-r} -\a r z^{-2} A^{-s}B^{-r-1}  = \Frac{1}{2} \sum_{k=-\infty}^{+\infty} k\,b_{s,r}^{(k)}(\alpha) z^{k-1} \ . 
\ee
This relation provides directly the recurrence relation
\be
\a s \, b_{s+1,r}^{(k-1)}(\a) -\a r \, b_{s,r+1}^{(k+1)}(\a) = k  b_{s,r}^{(k)}(\a) \ ,
\ee
but if we express $A^{-s-1}B^{-r}$ in terms of  $b_{s,r}^{(k)}(\a)$ and 
$A^{-s}B^{-r-1}$ in terms of  $b_{s,r+1}^{(k)}(\a)$, one obtains with  (\ref{eq.21})  the relation
\be
(1-\a^2) r b_{s,r+1}^{(k)}(\a) = (r-k) b_{s,r}^{(k)}(\a) + \a (s+k-1) b_{s,r}^{(k-1)}(\a) \ .
\label{eq.24}
\ee
Conversely, when  $A^{-s-1}B^{-r}$ is expressed in terms of  $b_{s+1,r}^{(k)}(\a)$ and 
$A^{-s}B^{-r-1}$ in term of  $b_{s,r}^{(k)}(\a)$, with (\ref{eq.22}), we have
\be
(1-\a^2) s b_{s+1,r}^{(k)}(\a) = (s+k) b_{s,r}^{(k)}(\a) + \a (r-k-1) b_{s,r}^{(k+1)}(\a) \ .
\label{eq.25}
\ee
Finally, when $A^{-s-1}B^{-r}$ and  $A^{-s}B^{-r-1}$ are expressed in terms of $b_{s,r}^{(k)}(\a)$,
we have
\be
(r-k-1) b_{s,r}^{(k+1)}(\a) =  (s+k-1) b_{s,r}^{(k-1)}(\a) +(\a(r-s-k) -\Frac{k}{\a})b_{s,r}^{(k)}(\a) \ .
\label{eq.26}
\ee
The five relations (\ref{eq.21}, \ref{eq.22}, \ref{eq.24}, \ref{eq.25}, \ref{eq.26}) allow then to 
express any generalized Laplace coefficient $b_{s,r}^{(k)}(\a)$ in terms of only two of them, namely
\be
 b_{s-[s],r-[r]}^{(0)}(\a) \qquad \hbox{and}\qquad   b_{s-[s],r-[r]}^{(1)}(\a)  \ .
\ee

In particular, all generalized Laplace coefficients involved in Eq. (\ref{eq.19}) can be expressed in terms 
of the two Laplace coefficients 
$
b_{[\gamma]-\gamma}^{(0)}(\a)$ and   $b_{[\gamma]-\gamma}^{(1)}(\a) $.

\section{Expressions in terms of true and eccentric anomaly}
The general Hansen coefficients  $X_k^{\g,m}$ can be expressed 
in simple form with respect to the Bessel functions $J_n(ke)$ and the generalized Laplace coefficients 
$b_{s,r}^{(k)}(\beta)$ (\ref{eq.19}), but 
although recurrence relations exist for both functions, they will not translate easily 
into recurrence relations for the Hansen coefficients  $X_k^{\g,m}$ when $k\neq 0$. On the opposite,
in the expression in terms of  the true anomaly  
\be
\left(\Frac{r}{a}\right)^{\g} \e^{imv} = \sum_{k=-\infty}^{+\infty} Y_k^{\g,m} \e^{ikv} \ ,
\label{eq.29}
\ee
or of the eccentric anomaly 
\be
\left(\Frac{r}{a}\right)^{\g} \e^{imv} = \sum_{k=-\infty}^{+\infty} Z_k^{\g,m} \e^{ikE}\ ,
\label{eq.30}
\ee
the Fourier coefficients $Y_k^{\g,m}$ and $Z_k^{\g,m}$ (see also
Brumberg, 1995) have very simple form in terms of Laplace and 
generalized Laplace coefficients. Indeed, from (\ref{eq.5}) and (\ref{eq.16a}), we have 
\be
Y_k^{\g,m} = \Frac{(-1)^{k-m}}{2}(1-e^2)^\g (1+\beta^2)^\g  b_{\g}^{(k-m)}(\beta) \ ;
\label{eq.31}
\ee
\be
Z_k^{\g,m} = \Frac{1}{2(1+\beta^2)^{\g}}b_{-\g+m,-\g-m}^{(k-m)}(\beta) \ .
\label{eq.32}
\ee
All the previous recurrence relations on Laplace and generalized Laplace coefficients can then 
be translated into 
recurrence relations on the  ($Y_k^{\g,m}$) and   ($Z_k^{\g,m}$) coefficients.

\section{Conclusion}
The computation of Hansen (1855), or the 
later computation of Hill (1875) and Tisserand (1889), are very similar to the present presentation, 
the novelty here being the use of Laplace, and generalized Laplace coefficients to express the Hansen coefficients in 
a simple form. This  explicits the fact that nowhere in the original demonstration 
of Hansen is requested the fact that $\g$ is an integer. This is particularly visible 
for the  computation of $X_0^{\g,m}$ that are very simple expressions of the  Laplace
coefficients $b_{\g+2}^{(m)}$.  As for the usual Laplace coefficients, 
recurrence relations  allow to express the generalized Laplace coefficients   $b_{s,r}^{(k)}$ in terms 
of only two of them,  $b_{s-[s],r-[r]}^{(0)}$ and $b_{s-[s],r-[r]}^{(1)}$, for example.  The 
Laplace and generalized Laplace coefficients are also introduced naturally  in the expression of 
$(r/a)^\g \exp(i m v)$ in Fourier series of the true and eccentric anomaly (29--32).

\section*{Acknowledgements}
The author is grateful to the referee for his suggestions and 
to J. Couetdic for a careful reading of the manuscript.

\end{article}
\end{document}